\documentclass[sort&compress]{elsarticle}
\pdfoutput=1
\usepackage[displaymath]{lineno}
\usepackage{multirow}
\usepackage[subrefformat=parens,labelformat=parens]{subfig}
\usepackage{natbib}
\begin{document}
\begin{frontmatter}
\title{LENDA, a Low Energy Neutron Detector Array for experiments with radioactive beams in inverse kinematics}

\author[NSCL,JINA]{G. Perdikakis\corref{cor1}}\ead{perdikak@nscl.msu.edu}
\author[NSCL,JINA]{M. Sasano}
\author[NSCL,JINA]{Sam M. Austin}
\author[NSCL] {D. Bazin}
\author[NSCL,JINA]{C. Caesar\fnref{fn1}}
\author[Hastings]{S. Cannon}
\author[NSCL,JINA]{J. M. Deaven}
\author[NSCL,JINA]{H. J. Doster}
\author[NSCL,JINA]{C. J. Guess \fnref{fn5}}
\author[NSCL,JINA]{G. W. Hitt \fnref{fn4}}
\author[NSCL]{J. Marks}
\author[NSCL,JINA] {R. Meharchand \fnref{fn2}}
\author[NSCL]{D. T. Nguyen}
\author[MSU]{D. Peterman}
\author[NSCL,JINA]{A. Prinke}
\author[NSCL,JINA]{M. Scott}
\author[NSCL]{Y. Shimbara \fnref{fn3}} 
\author[MichTech] {K. Thorne}
\author[NSCL,JINA]{L. Valdez}
\author[NSCL,JINA,MSU]{and R. G. T. Zegers}

\address[NSCL]{National Superconducting Cyclotron Laboratory, Michigan State University, East Lansing, MI 48824, USA}
\address[JINA]{Joint Institute of Nuclear Astrophysics, Michigan State University, East Lansing, MI 48824, USA}
\address[MSU]{Department of Physics and Astronomy, Michigan State University, East Lansing, MI 48824, USA}
\address[MichTech]{Michigan Technological University, Houghton, MI 49931-1295, USA}
\address[Hastings]{Hastings College, Hastings, NE 68901, USA}

\cortext[cor1]{Corresponding author}
\fntext[fn1]{Present address: GSI Helmholtzzentrum f\"ur Schwerionenforschung GmbH, 64291 Darmstadt, Germany}
\fntext[fn5]{Present address: Department of Physics and Applied Physics, University of Massachusetts Lowell, Lowell, MA 01854}
\fntext[fn4]{Present address: Department of Applied Mathematics and Sciences, Khalifa University of Science, Technology, and Research, P.O. Box 127788, Abu Dhabi, UAE}
\fntext[fn2]{Present address: Los Alamos National Laboratory, Los Alamos, New Mexico 87545, USA}
\fntext[fn3]{Present address: Graduate School for Science and Technology, Niigata University, Niigata 950-2181, Japan}

\begin{abstract}
The Low Energy Neutron Detector Array (LENDA) is a neutron time-of-flight (TOF) spectrometer developed at the National Superconducting Cyclotron Laboratory (NSCL) for use in inverse kinematics experiments with rare isotope beams. Its design has been motivated by the need to study the spin-isospin response of unstable nuclei using ($p,n$) charge-exchange reactions at intermediate energies ( $ > $ 100 MeV/u). It can be used, however, for any reaction study that involves emission of low energy neutrons (150 keV - 10 MeV). The array consists of 24 plastic scintillator bars and is capable of registering the recoiling neutron energy and angle with high detection efficiency. The neutron energy is determined by the time-of-flight technique, while the position of interaction is deduced using the timing and energy information from the two photomultipliers of each bar. A simple test setup utilizing radioactive sources has been used to characterize the array. Results of test measurements are compared with simulations. A neutron energy threshold of $<$ 150 keV, an intrinsic time (position) resolution of $\sim$ 400 ps ($\sim$ 6 cm) and an efficiency $ > $ 20 \% for neutrons below 4 MeV have been obtained.
\end{abstract}
\begin{keyword}
neutron detector \sep neutron Time-of-Flight \sep low energy neutron detector \sep inverse kinematics \sep ($p,n$) charge-exchange
\end{keyword}
\end{frontmatter}
\section{Introduction}
A high-efficiency position-sensitive scintillator array for the detection of low energy neutrons has been developed at the National Superconducting Cyclotron Laboratory (NSCL). The Low Energy Neutron Detector Array (LENDA) was built to study charge-exchange (CE) ($p$,$n$) reactions. These reactions have been successfully used in the past to study the spin-isospin response of stable nuclei \cite{Osterfeld92,harakeh01}. LENDA makes possible the extension of such studies to unstable nuclei using the ($p$,$n$) CE reaction in inverse kinematics \cite{Perdikakis09a}. Moreover, LENDA can be utilized in any nuclear reaction study that involves production of low-energy neutrons, for example, for proton-transfer ($d$,$n$) reactions in inverse kinematics and beta-delayed neutron emission experiments. In such investigations, the detection of the slow neutrons with good efficiency as well as energy and angle resolution is essential. LENDA addresses this need by providing fast timing ($\sim$400 ps timing resolution), reasonable position resolution ($\sim$ 6 cm ) and a low neutron detection threshold of 150 keV.\\
For the characterization of the array, a simple test setup utilizing neutrons emitted by a $ ^{252} $Cf radioactive source and photons from a $^{22}$Na source has been used. In this work, the results of the characterization are presented. The setup is described in section \ref{sec:Describe LENDA}. The electronics and data acquisition system (DAQ) is presented in section \ref{sec:DAQ}, a brief description of the data reduction process in section \ref{sec:DataReduc}; the efficiency, timing and position resolution of LENDA in sections \ref{sec:Efficiency}, \ref{sec:Timing} and \ref{sec:Position} respectively. The conclusion follows in section \ref{sec:Conclusion}.
\section{Description of LENDA}
\label{sec:Describe LENDA}
LENDA is an array of 24 neutron detectors. 
Each detector module is a type BC-408 \cite{SaG11} plastic scintillator bar with dimensions of $300 \times45 \times 25$ mm.
A Hamamatsu H6410 photomultiplier (PMT) assembly with a photo-electron gain of the order of 10$^{7}$ is used at each end of the bar to detect the scintillation light. In each module, to ensure optimum transmission of light to the photocathode, the PMTs are directly coupled to the scintillator using optical epoxy. The scintillator bar is wrapped with one layer of white nitrocellulose membrane filter paper from Advantec \cite{Adv11}, a layer of aluminum foil and finally black insulating tape to ensure proper light propagation through the bar as well as light-tightness. The Advantec film has comparable performance to other high reflectivity scintillator wrapping materials available commercially (Gore DRP \cite{Gore11}, 3M Vikuity \cite{3MVik11}) and is easier to manipulate while wrapping the LENDA scintillators.\\
\begin{figure}[h]
\centering
\includegraphics[width=0.4\linewidth]{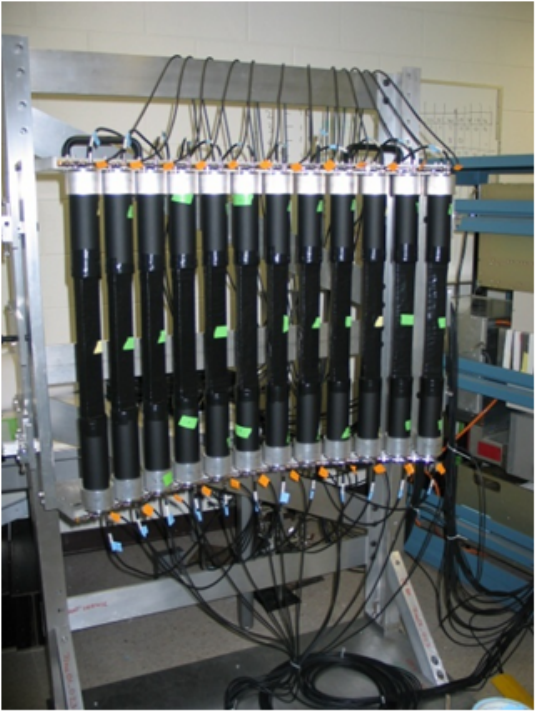}
\caption{One half of the LENDA array (12 detector modules) mounted on its stand. In this configuration, optimized for charge-exchange (p,n) experiments in inverse kinematics, the bars are arranged at a radius of 1 m from the target position. One stand is placed on each side of the beam line and the array covers roughly 45 degrees of scattering angle in the laboratory frame.}
\label{fig:array}
\end{figure}
Figure \ref{fig:array} shows a picture of one half of the array. Each bar is mounted so that the position-sensitive direction of the bar is vertical, and the shortest side of the bar is parallel to the neutron flight path. At a distance of 1 m from a neutron source, one array covers a solid angle of 0.16 sr in total for a scattering-angle coverage of about 45 degrees.\\
By combining the time and pulse-height information from the PMTs, the timing of a neutron hit, the corresponding scintillation light output, and the neutron's hit position along the longest side of each LENDA bar can be determined. During standard Time-of-Flight (TOF) operation mode, LENDA has to be used with an external detector that provides a time reference signal. In a typical experimental setup where LENDA is used in combination with the S800 spectrometer of NSCL \cite{Baz03}, that signal is generated by charged particles hitting a diamond detector \cite{Stolz06} at the object of the spectrometer.
\section{Data Acquisition system}
\label{sec:DAQ}
\begin{figure}[h]
\includegraphics[width=0.9\linewidth]{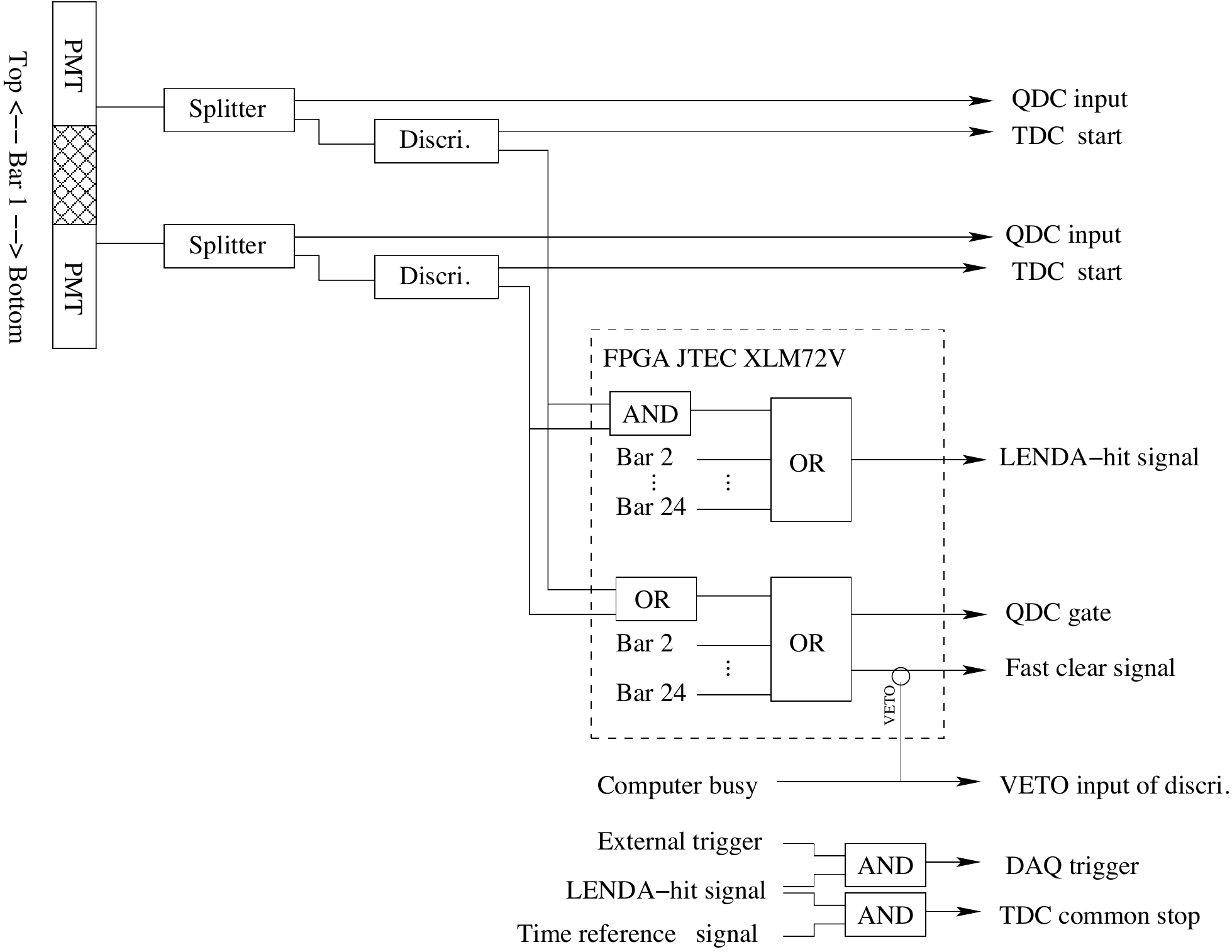}
\caption{A schematic diagram of the DAQ system for LENDA.}
\label{fig:daq}
\end{figure}

A common-stop trigger logic is employed for the time-of-flight measurement. The trigger logic is realized using a Xilinx Virtex-II FPGA chip \cite{Xil11} implemented in a VME-based JTEC XLM72V FPGA module \cite{JTEC11}. Two 32-channel CAEN V792 charge-to-digital converters (QDCs) \cite{Caen11} are used to digitize the charge of the PMT signals. The time information for each event is digitized by two 32-channel CAEN V775 time-to-digital converters (TDCs). 
The PMTs are powered by ISEG EHS F030n and 8030n high voltage power supplies. Figure \ref{fig:daq} shows a schematic diagram of the LENDA DAQ. To achieve the highest possible signal-to-noise ratio from the PMTs and thus maintain a low threshold of neutron detection, the PMT voltage is set to be as close as possible to the maximum value of -2700 Volts permitted by the manufacturer. Some voltage variation (50 - 100~Volts in most cases) between individual PMTs is employed to obtain a rough gain-matching of the signals. The anode signal of each PMT is attenuated by a factor of 0.3 using a custom-built 16-channel, two-way splitter-attenuator module, before it is used as an input to the DAQ electronics. This is done to match the signal's pulse-height to the dynamic range of the digitizers and discriminators for the neutron energies of interest.
For each PMT, one of the split anode signals, after being cable-delayed by $\sim 100$ ns, is sent to the QDC  to digitize its charge. The other split signal is sent to a Phillips 7106 leading edge discriminator to generate a logic signal that indicates the PMT has fired (PMT-event signal).
The PMT-event signal is sent to the TDC and serves as a start-signal for the TOF digitization of that event. 
It is also used as one of the input signals for the FPGA trigger-logic. One more input signal for the logic is required to identify a ``good" event. That signal is provided by an external trigger-detector that is experiment-specific. Five logic signals appear at the output of the FPGA as a result of the trigger processing:
\begin{itemize}
\item A LENDA-hit signal, which signifies that at least one LENDA bar is hit.
This is generated by the coincidence of the PMT-event signals from the two ends of each LENDA bar.
An OR between those coincidences from all LENDA bars gives the LENDA-hit signal.
\item A QDC gate signal provided by the OR of all PMT-event signals.
\item A FAST CLEAR signal generated by the same OR, with a delay of a few hundred ns. This clears the digitizers and aborts the processing of the signal when the DAQ is not triggered by a good event.
\item A VETO signal that eliminates the FAST CLEAR signal whenever a coincidence is realized (DAQ trigger) between the external trigger and the LENDA-hit signal. This allows ``good" data to be digitized successfully.
\item A COMMON STOP signal that causes the TDC to digitize the timing information. This is, in terms of timing, the same as the DAQ trigger signal.
\end{itemize} 
The FAST CLEAR scheme that allows events to be selected after the QDC integration has an advantage. The starting time of the QDC integration can be set independently of the timing of the trigger signal. This reduces the amount of delay required to place the detector signal inside the QDC gate.
\section{Data Reduction}
\label{sec:DataReduc}
\begin{figure}
\centering
\includegraphics[width=0.6\linewidth]{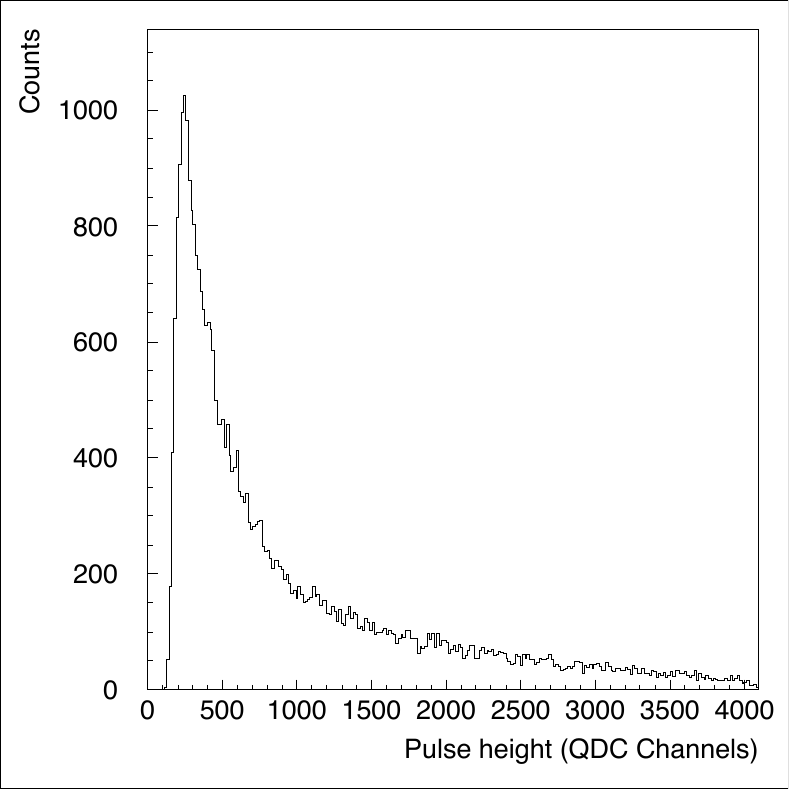}
\caption{QDC spectrum of neutrons emitted by a $^{252}$Cf source at a distance of 1m from the detector.}
\label{fig:QDC_Spectrum}
\end{figure}

Neutron TOF, hit position along the longest side of a scintillator bar, 
and light output in the scintillator 
are obtained by
combining the timing of signals from the PMTs at the top and bottom ends ($t_T$ and $t_B$) of the scintillator and the signal amplitudes ($q_T$ and $q_B$) recorded in the TDC and QDC modules, respectively. Herein, we briefly describe the definition and derivation of these parameters. This analysis is similar as described by other authors~\cite{Pal01, Wak05}.\\
Owing to quenching effects and the propagation time of scintillation light from the hit position in the scintillator to the PMTs, 
the signal timing and amplitude are both asymmetric between the top and bottom PMTs. This asymmetry is correlated with the hit position. 
Two position-dependent parameters $x_T$ and $x_Q$
are derived from the difference between $t_T$ and $t_B$, and between $q_T$ and $q_B$ using the relations,
\begin{eqnarray}
  x_T &\propto& T_{diff}= t_T - t_B, \label{eq:x_T}\\
  x_Q &\propto& cogQ = \frac{q_T - q_B}{q_T + q_B}.\label{eq:x_Q}
\end{eqnarray}
The proportionality between $x_{Q}$ and $cogQ$ in equation \ref{eq:x_Q} is a valid approximation in the case of LENDA: The length of the scintillator bars (30~cm) is much smaller than the light attenuation length of 210~cm for the BC-408 material \citep{SaG11}; then, the exponential dependence of $q_T$ and $q_B$ on hit position can be approximated by its Taylor expansion leading to a linear relationship.\\   
The neutron TOF ($t$) and light output ($l$), are derived from an average of $t_T$ and $t_B$, and from $q_T$ and $q_B$, respectively, as
\begin{eqnarray}
  t&=&\frac{t_T+t_B}{2}-t_{TR},\\
  l&=&\sqrt{q_T \times q_B},
\end{eqnarray}
where $t_{TR}$ is the time reference (TR) for TOF provided by the external detector. The $q_T$ and $q_B$ values are converted from QDC channels to light output units of electron equivalent energy (MeV$_{ee}$), by assuming a linear relation between the QDC values and the light output. This approach has been shown to reproduce quite accurately the experimentally measured light output for organic scintillators \cite{Kornilov09}. A typical QDC spectrum for neutrons emitted from a $^{252}$Cf source is shown in figure \ref{fig:QDC_Spectrum}.
The relation connecting QDC values to light output is calibrated using the 59~keV $\gamma$-ray of $^{241}$Am and the Compton edge of the 662-keV $\gamma$-ray from a $^{137}$Cs source.\\
\section{Neutron spectrum and efficiency with $^{252}$ Cf}
\label{sec:Efficiency}
Time-of-flight spectra of neutrons from a $^{252}$Cf fission source  were measured by LENDA to determine its neutron detection efficiency. The kinetic energy of fission neutrons from $^{252}$Cf ranges approximately from 100~keV to 10~MeV according to the IAEA evaluated neutron energy distribution \cite{Man87a,Man89a}. This distribution was used to reliably determine the energy dependence of the efficiency curve.
\begin{figure}[h]
\centering
\includegraphics[width=0.9\linewidth]{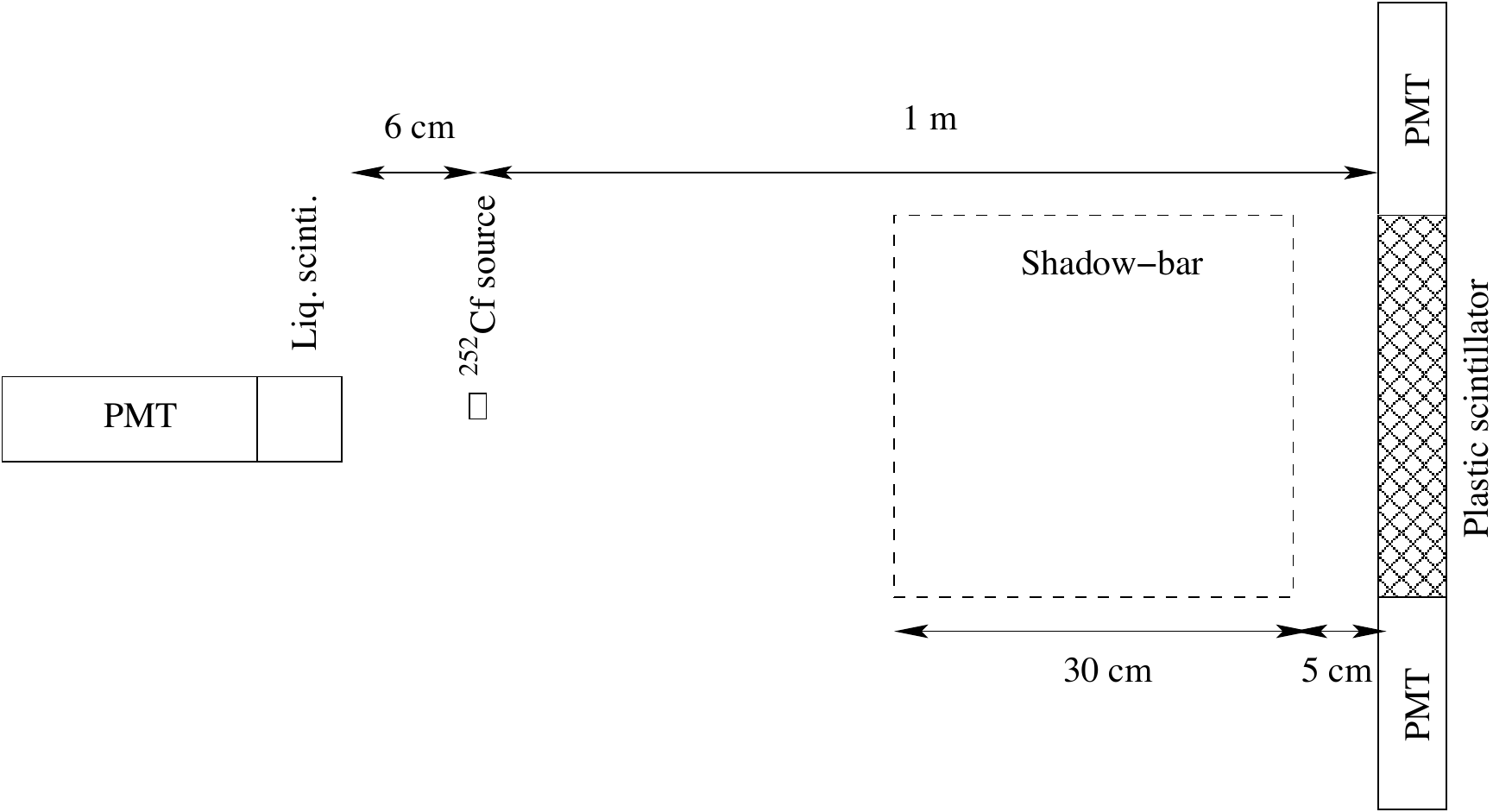}
\caption{A schematic view of the experimental setup used in the measurement of a $^{252}$Cf neutron spectrum. The $^{252}$Cf fission source was placed at a distance of 1~m from the LENDA bar. Prompt $\gamma$-rays from fission were detected in a liquid scintillator. Neutron-$\gamma$ ray pulse shape discrimination was used to separate them from prompt neutrons.
The $\gamma$-rays identified in the liquid scintillator served as the time reference for the neutron TOF. The shadow bar was used to determine the contribution from background events due to wall-scattered neutrons. See the text for details.}
\label{fig:eff_setup}
\end{figure}\\
In the experimental setup (shown schematically in figure ~\ref{fig:eff_setup}), the $^{252}$Cf source was placed at a distance of 1~m from the LENDA bars. The TR signal for the neutron TOF was established by detecting prompt $\gamma$-rays from fission in a $2^{\prime\prime}\times2^{\prime\prime}$ cylindrical NE-213 type liquid scintillator manufactured by ELJEN (EJ-301) \cite{Elj11a}. This detector was placed at a distance of 6~cm from the source. Pulse shape neutron-$\gamma$ discrimination was employed in the TR scintillator for selecting only $\gamma$-ray events. In all TOF spectra a correction for the ``walk" of the leading edge discriminator was applied in the off-line analysis. 
\begin{figure}[h]
\centering
\includegraphics[width=0.6\linewidth]{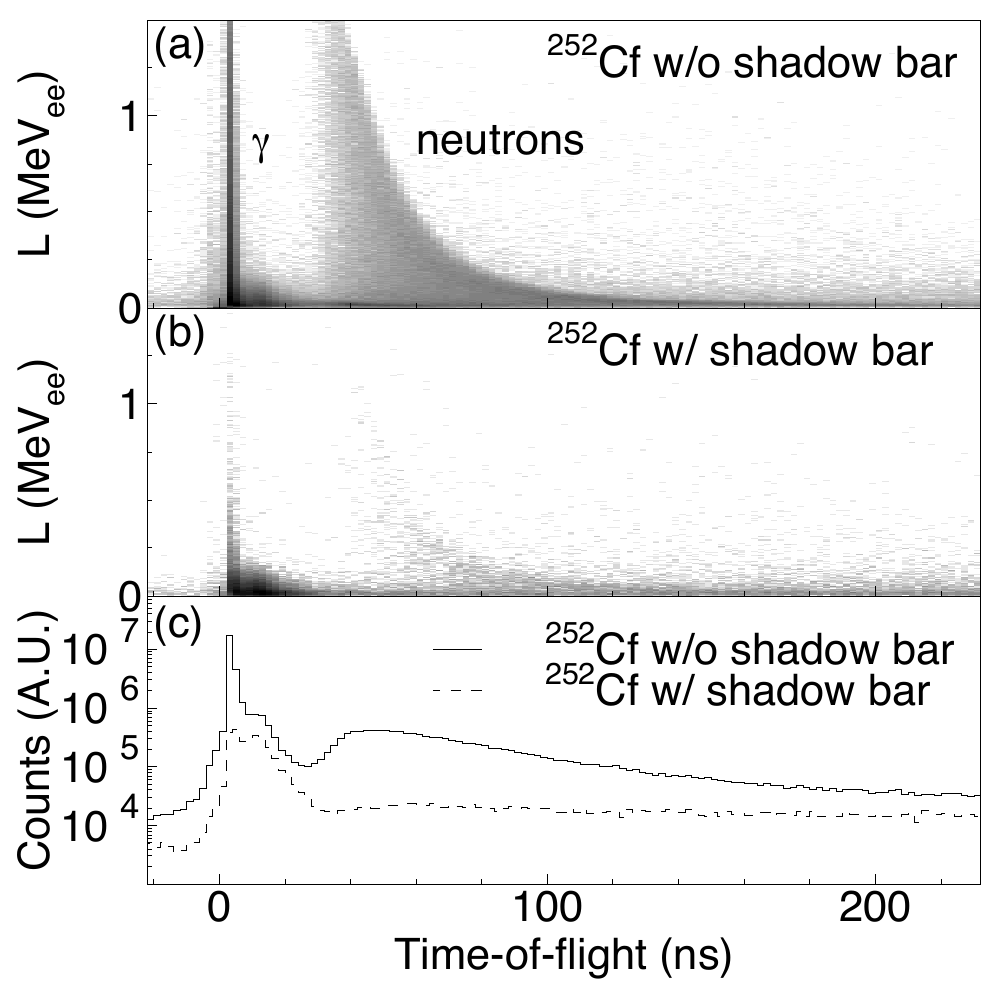}
\caption{Response of LENDA to prompt neutrons and $\gamma$-rays from $^{252}$Cf fission events shown as a function of the TOF and light output in the scintillator bar with (a) and without (b) a shadow bar. The projection onto the TOF axis of (a) and (b) is shown in (c).}
\label{fig:PIDToF}
\end{figure}
Figure~\ref{fig:PIDToF}(a) shows a 2-dimensional spectrum of light output versus TOF for $\gamma$ rays and neutrons detected in a LENDA bar.  The continuous curve in figure~\ref{fig:PIDToF}(c) shows the projection of the 2-dimensional spectrum on its horizontal (TOF) axis. In figure~\ref{fig:PIDToF}(a), the events included in the sharp peak at $t\sim3$~ns are due to $\gamma-\gamma$ coincidences between the LENDA and TR detectors. The events that lie on the right side of the spectrum at $t > 20$~ns correspond to $\gamma-n$ coincidence events (neutrons detected in LENDA). Events appearing between those two regions are due to $\gamma$ rays from the source, reaching LENDA bars after being scattered from the surrounding walls. Any events in the region of $t < 3$~ns on the left of the $\gamma$ ray peak come from neutrons leaking through the n-$\gamma$ discrimination gate that were being detected in the liquid scintillator in coincidence with $\gamma$ rays  in LENDA.\\
Since increasing TOF corresponds to lower neutron kinetic energies, the maximum detected light output in a plastic scintillator like LENDA is varying as a function of TOF. The characteristic light output curve corresponding to full energy deposition by neutron events, becomes practically zero for TOF values of $t \sim 150$~ns.  The TOF for this minimum amount of light output corresponds to the smallest kinetic energy that can be detected by LENDA, about 150~keV. It is defined by the lowest threshold setting of the DAQ discriminator module and the maximum gain of the PMTs.\\
The events lying above the light output curve are background events, which are mostly due to neutrons and gammas indirectly reaching LENDA after scattering by the surrounding walls or objects. For these events, TOF is not correlated with the deposited energy. In the offline analysis, the contribution from this background can be reduced by eliminating the events in the region above the light output curve (light-output cut).\\
The remaining background contribution was directly measured by inserting a copper shadow bar to block neutrons directly coming from the source to LENDA as shown in figure~\ref{fig:eff_setup}. The dimensions of the shadow bar were selected so that they match the size of the scintillator bar. It was thick enough (30~cm) to attenuate neutrons with kinetic energy up to 10~MeV, by a factor larger than 600. The effect of the shadow bar blocking part of the background neutron flux was calculated assuming an isotropic background neutron distribution. The shadow bar concealed only 5\%  of the total solid angle subtended by LENDA and the systematic error induced in the efficiency calculation was estimated to be of the order of 1\%. The live time of the measurement was used to normalize the background runs to those without the neutron blocker. The spectrum measured with the shadow bar is shown in figure~\ref{fig:PIDToF}(b), where most of the neutron events are due to scattering from walls. The dashed curve in figure~\ref{fig:PIDToF}(c) corresponds to the projection of this 2-dimensional spectrum on the TOF axis.\\
To determine the absolute value of the efficiency, the $^{252}$Cf neutron yield distribution of Refs.~\cite{Man87a,Man89a} was normalized using a yield reference (YR) detector of the same type and size as the TR one. The YR detector was placed at a distance of 40~cm from the source. Its efficiency was simulated using the computer code NEFF7~\cite{Dic98a} and found to vary between 30 and 40\% for neutrons with energies in the region of 1 to 3~MeV. According to previous studies using fission fragment detectors~\cite{cub89}, the results of efficiency simulations using the NEFF7 code are known to be reliable within an error of 5\% for this type of detector.\\
After subtracting the contribution from the background, the number of detected neutrons, $N(E_n)$, as a function of neutron energy, $E_n$, are compared to the neutron yield from the $^{252}$Cf source, $Y(E_n)$ using the relation,
\begin{equation}
\centering
  N(E_n) = Y(E_n) \times \epsilon(E_n) \times \Delta\Omega,
\label{eq:eff1} 
\end{equation}
where $\Delta\Omega$ is the solid angle of the LENDA bar precisely calculated from the alignment information of the detector and source, $Y(E_n)$ is the standard shape of neutron energy distribution taken from Refs.~\cite{Man87a,Man89a} and normalized using the YR detector, and the $\epsilon(E_n)$ is the neutron-detection efficiency of the LENDA bar as a function of $E_n$.
\begin{figure}[h]
\centering
\includegraphics[width=0.6\linewidth]{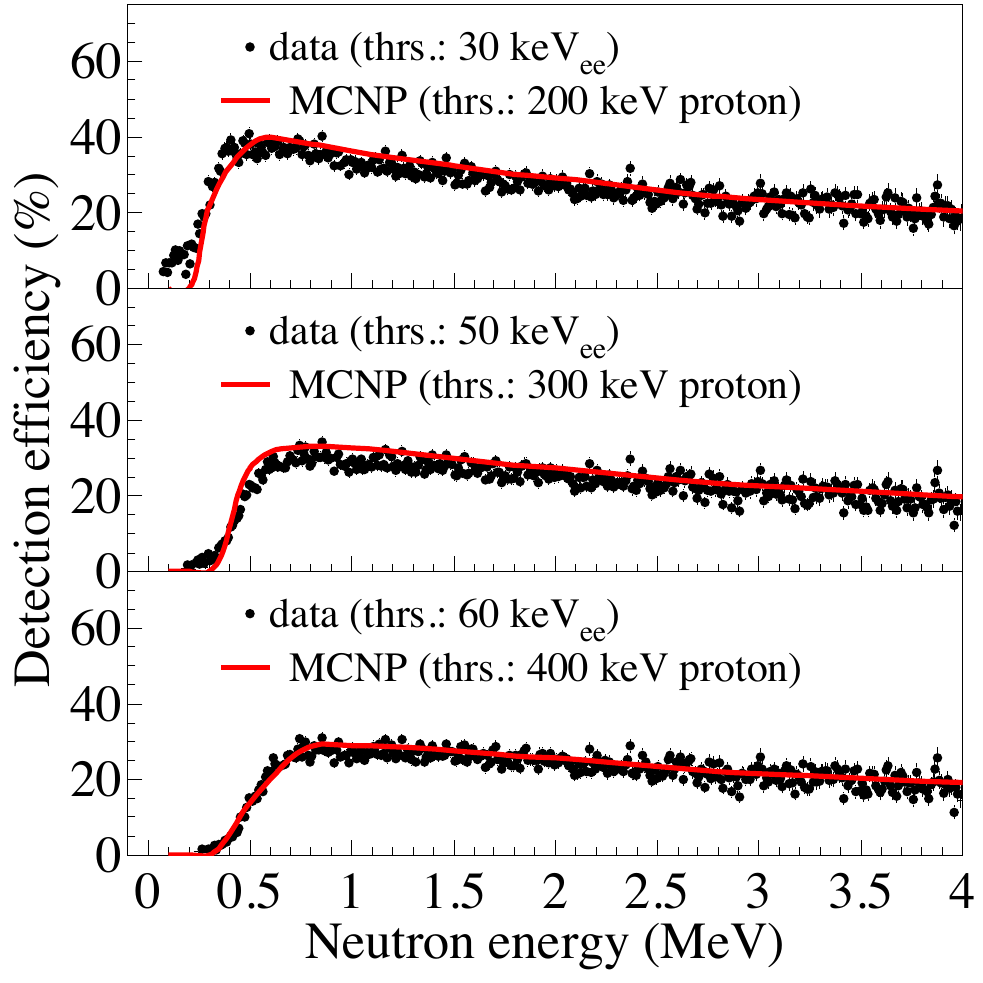}
\caption{Experimental neutron detection efficiencies of a LENDA bar (dots) as a function of neutron energy in comparison with the curve simulated by MCNPX (curves). The error bars (shown where they are larger than the points) indicate statistical uncertainties. See the text for details.}
\label{fig:eff}
\end{figure}
The obtained efficiency curves are shown with dots in Fig.~\ref{fig:eff}, where three threshold levels of 30, 50, and 60 keV$_{ee}$ on the light output were applied in the offline analysis. The results of the simulation of the setup using Monte Carlo N-Particle Transport Code (MCNPX)~\cite{MCNPX02a} are shown for comparison. The thresholds for the recoiling proton energy in the simulations were set to be 200, 300, and 400 keV, respectively, to match the light-output threshold in the experimental data. The simulated distributions reproduce fairly well the trend and magnitude of the measured efficiency for the three values of the threshold. Any small deviations are probably due to ambiguities in background subtraction.
\section{Timing resolution}
\label{sec:Timing}
The timing resolution of LENDA was determined using the two correlated 511 keV photons from the positron annihilation of $ ^{22} $Na. Due to compton-scattering (the dominant mechanism of photon detection in BC-408 at these energies) photons of 511~keV produce a continuum of low-energy events which are equivalent (in terms of pulse height) to the proton recoils produced in the scintillator by 2.3~MeV neutrons. The timing response of the BC-408 plastic scintillator to gammas and neutrons  is therefore the same (same pulse shape) with the pulse-height of the compton edge of 511~kev photons corresponding to a neutron energy of 2.3~MeV.\\
\begin{figure}[h!]
\centering
\includegraphics[width=0.5\linewidth]{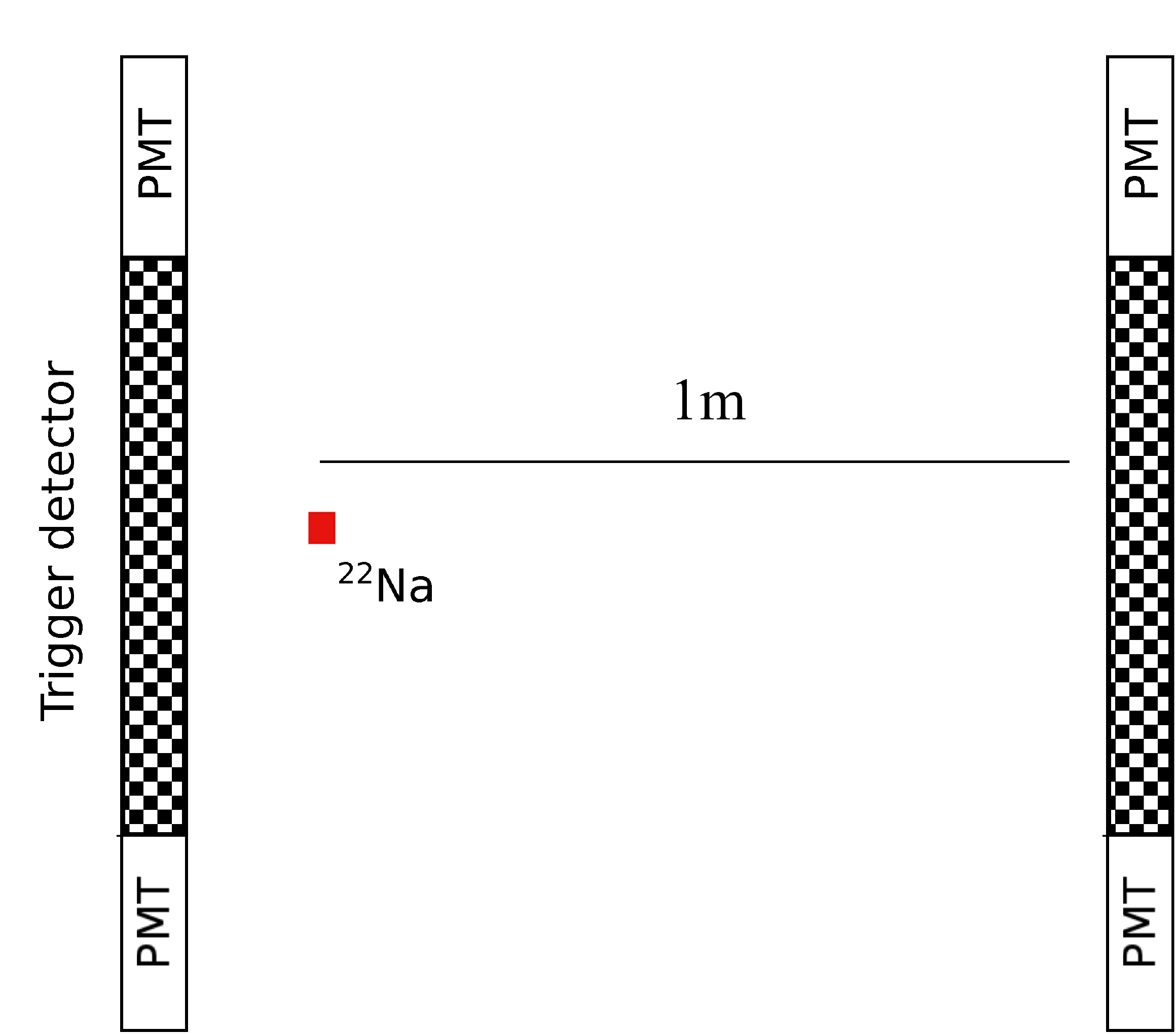}
\caption{Schematic depiction of the setup used to determine the timing resolution of LENDA bars. Photons from a $^{22}$Na positron source were used. In this setup, one of the LENDA bars was functioning as the trigger detector providing the time reference signal for the DAQ. }
\label{fig:setup_resol_22Na}
\end{figure}
For the timing resolution measurements, two LENDA bars were placed  opposite to each other so that the source would be at a distance of 1 m from one of the bars, as shown in figure \ref{fig:setup_resol_22Na}. One of the LENDA bars (closest to the source) served as a trigger detector while the other registered the photon TOF.
\begin{figure}[h!]
\centering
\includegraphics[scale=1.0]{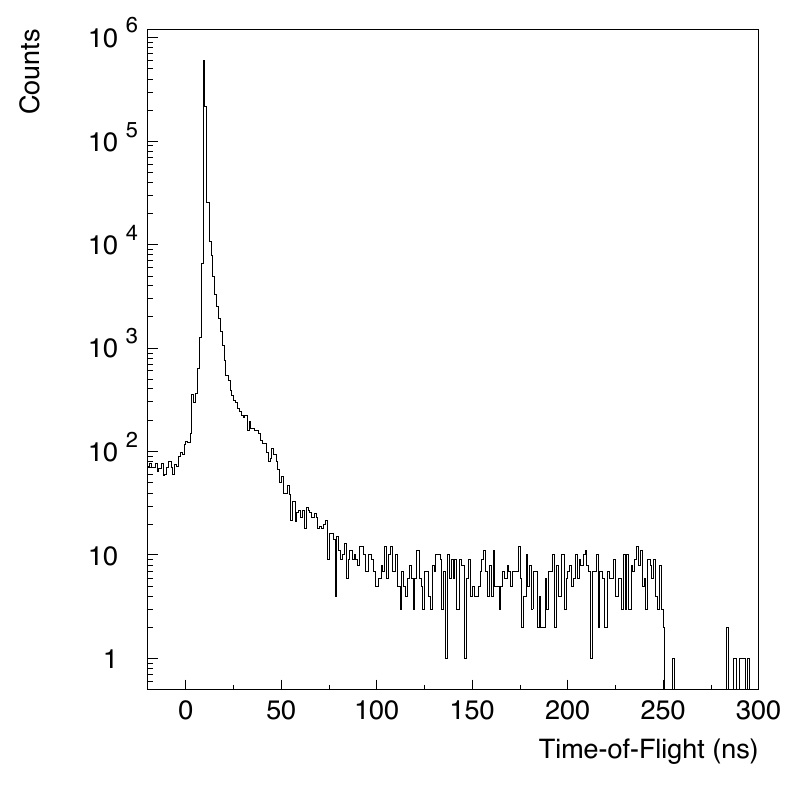}
\caption{Spectrum representing the timing resolution obtained with LENDA using a $ ^{22}$Na positron emitting source in the arrangement shown in figure \ref{fig:setup_resol_22Na}. The peak corresponds to $\gamma$-$\gamma$ photon coincidence events from positron annihilation that are detected by the two LENDA bars. The width of the peak corresponds to a timing resolution of $\approx$420 ps. }
\label{fig:Resol}
\end{figure}
\begin{figure}[h]
\centering
\includegraphics[width=0.7\linewidth]{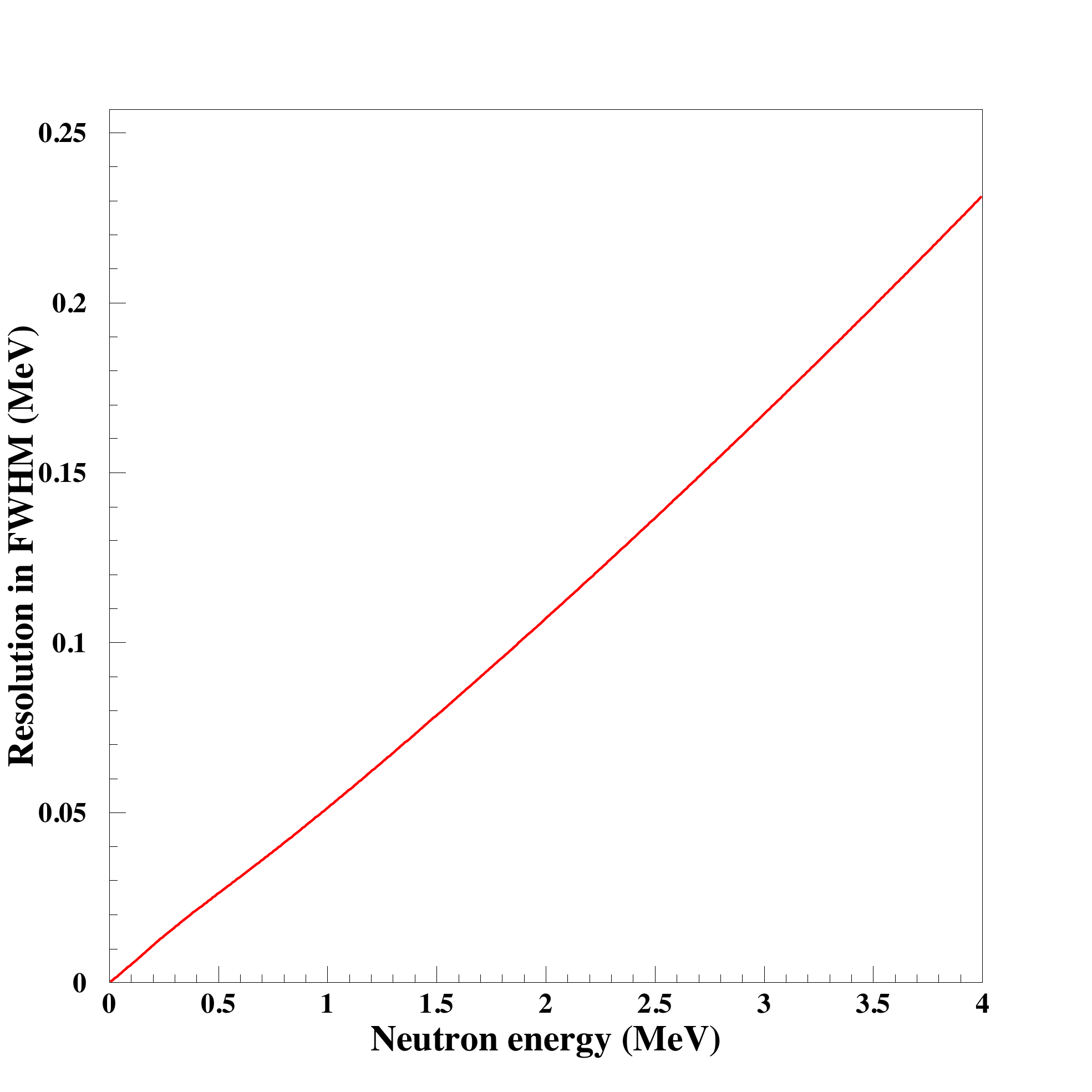}
\caption{Plot of the energy resolution achievable by LENDA for neutron energies in the range of 0 - 4 MeV, based on measured timing resolution for photons of 511~keV. }
\label{fig:de_resol}
\end{figure}
In the time spectrum of figure \ref{fig:Resol} a prominent peak corresponding to $\gamma$-$\gamma$ coincidences is observed.
The width of the photon peak in figure \ref{fig:Resol} is roughly 600 ps FWHM which corresponds to an intrinsic resolution of 600 ps$ / \sqrt{2} $ $ \approx $ 420 ps for each bar. This timing resolution is nearly constant for light outputs above 200~keV$_{ee}$ and degrades fast for smaller light outputs. As shown in figure \ref{fig:de_resol}, this timing resolution corresponds to an almost constant 5-6\% $ \Delta E/E $ resolution for neutron energies below 4 MeV when detected by LENDA at a distance of 1 m. The degradation of timing resolution for light outputs below 200~keV$_{ee}$ causes to the energy resolution only a small deviation from linearity for low energy neutrons (below roughly 1.3~MeV). In an case this degradation of the timing resolution does not significantly affect the energy resolution of the detector at the corresponding energies.
\section{Position Resolution}
\label{sec:Position}
\begin{figure}[h]
\centering
\includegraphics[width=0.9\linewidth]{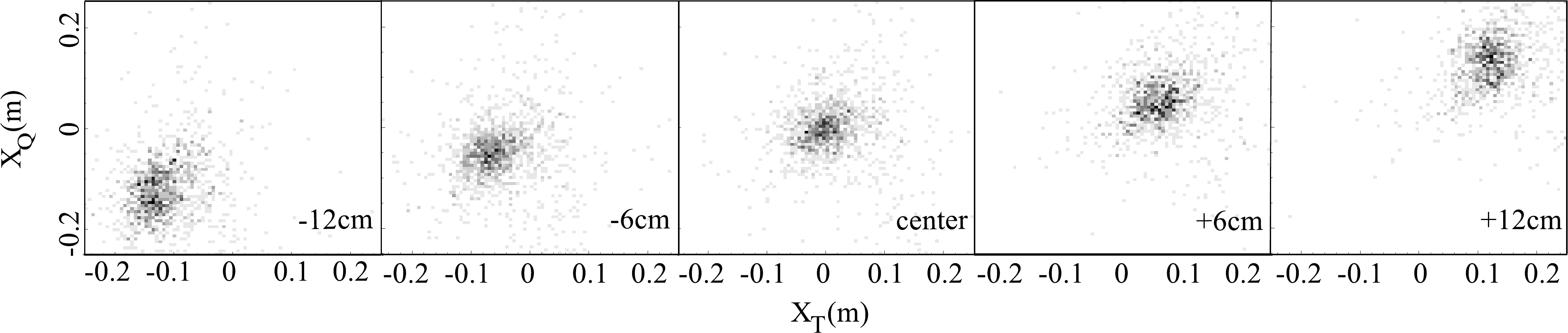}
\caption{Events from a source attached on the long side of a LENDA bar generate a locus in the 2-D spectrum of time difference versus charge asymmetry. In this figure the two axes are labeled X$_{T}$ and X$_{Q}$ respectively and calibrated in units of distance (m). The locus of events changes with source location along the scintillator bar, revealing the position sensitivity of the detector. In each figure the distance of the source from the middle point (center) of the bar is displayed as well.}
\label{fig:position_slideshow}
\end{figure}

\begin{figure}[h]
\centering
\includegraphics[width=0.4\linewidth]{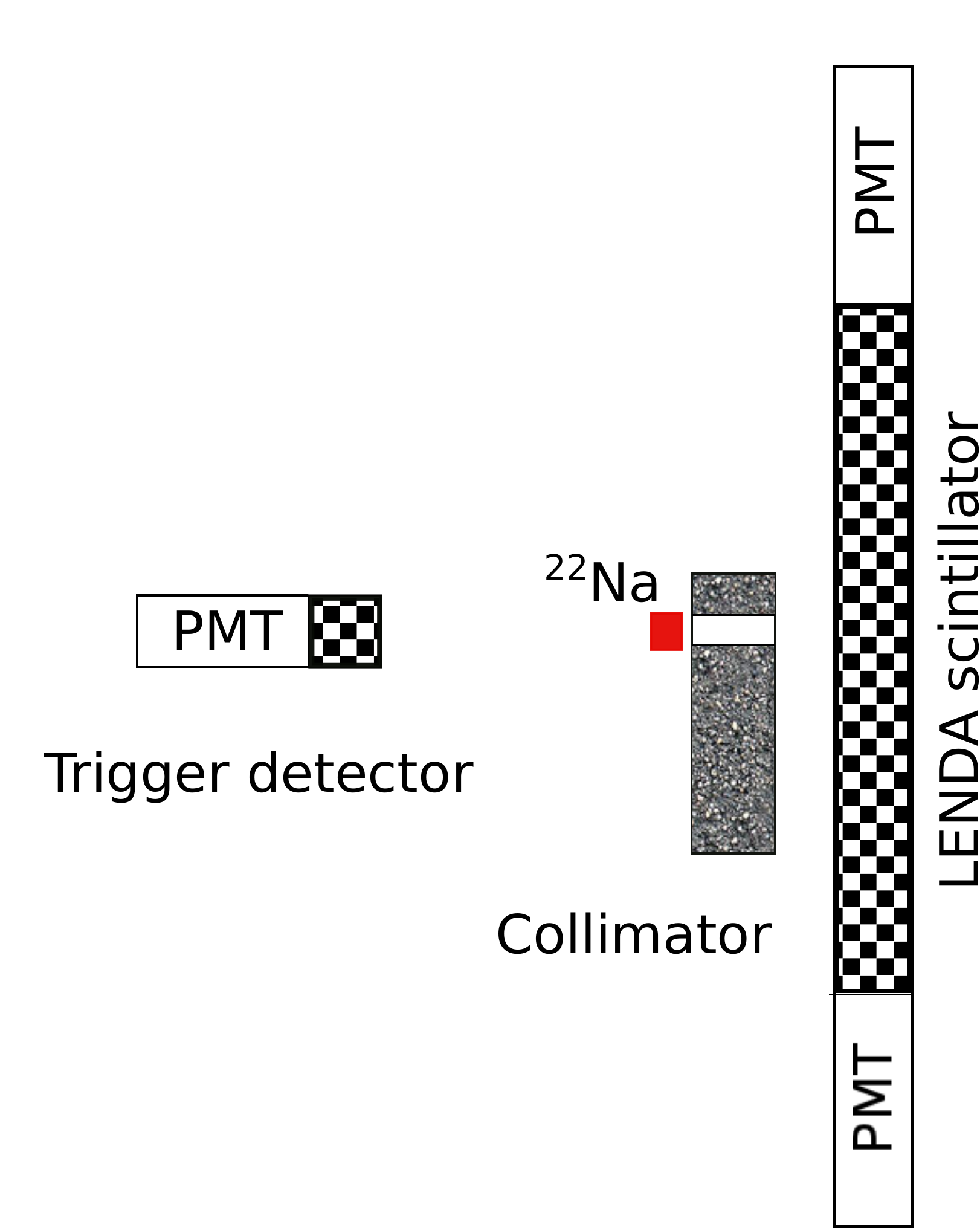}
\caption{Picture of the setup used for position calibration}
\label{fig:position_setup}
\end{figure}

\begin{figure}
\centering
\includegraphics[width=0.6\linewidth]{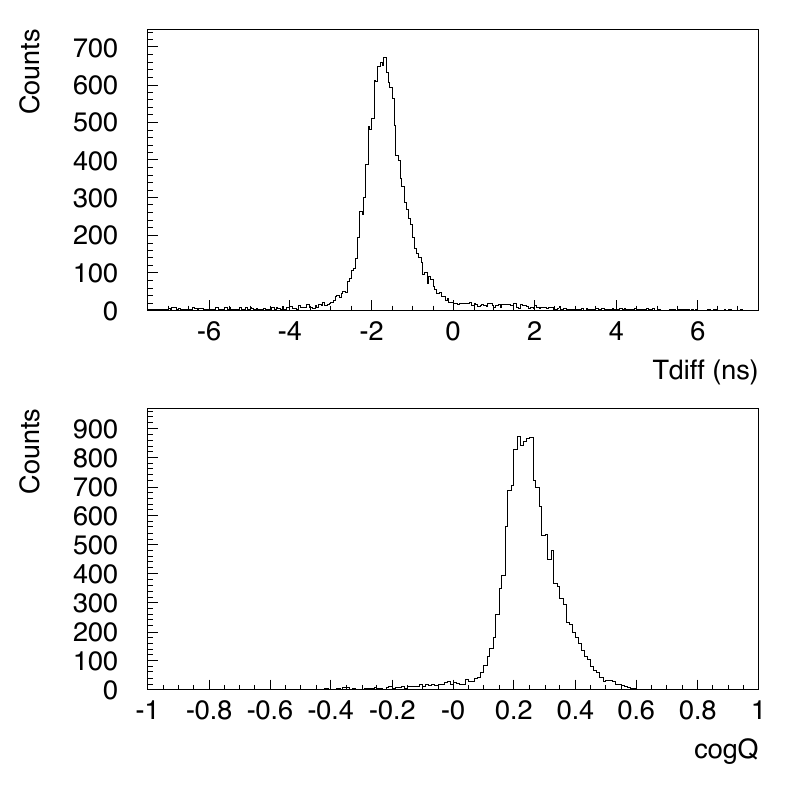}
\caption{Plot of the T$_{diff}$ (top figure) and cogQ (bottom figure) distributions that correspond to one location of the collimated source along a LENDA bar.}
\label{fig:pos_distributions}
\end{figure}

\begin{figure}[h]
\centering
\subfloat[Tdiff vs X]{\label{fig:TdiffvsX}\includegraphics[width=0.5\linewidth]{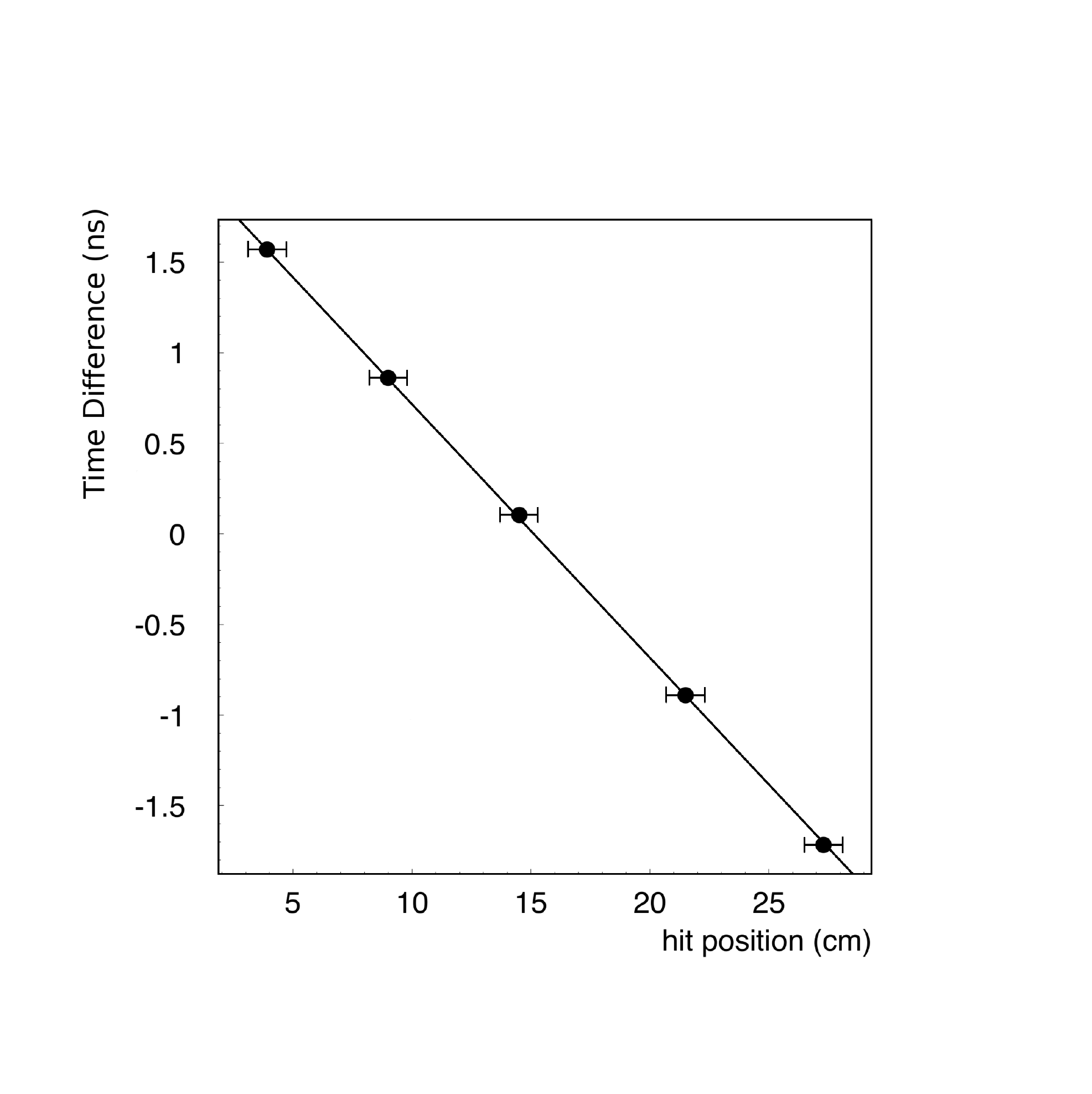}}\hspace{-2em}
\subfloat[cogQ vs X]{\label{fig:CogvsX}\includegraphics[width=0.5\linewidth]{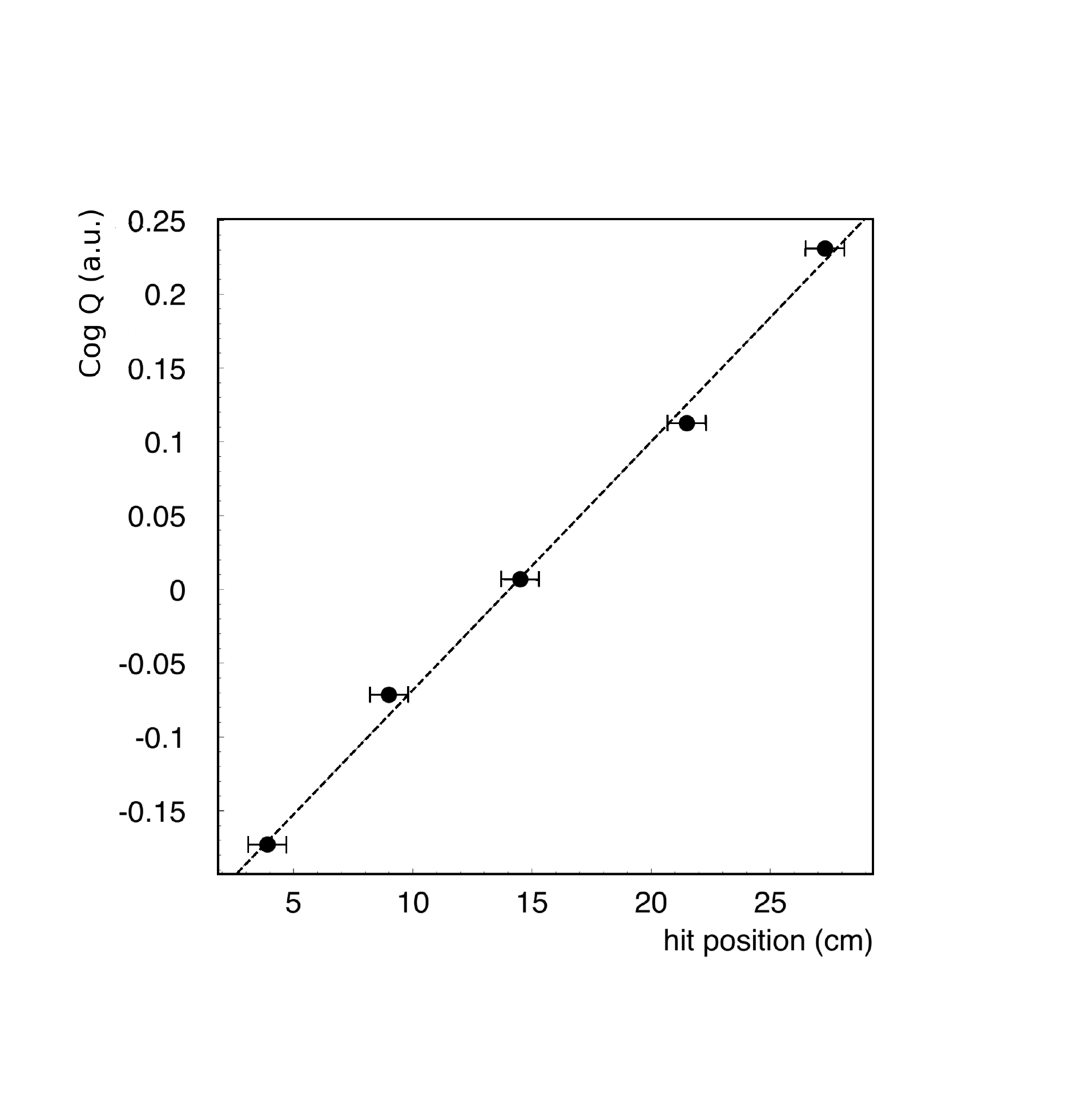}}
\caption{Position dependence of (a) the time difference and (b) of the charge center of gravity between the two PMTs of a LENDA bar. The straight lines in the graphs correspond to a linear fit of the data. The error in the horizontal axis corresponds to the diameter of the aperture used to collimate the source.}
\end{figure}

\begin{figure}[h]
\centering
\includegraphics[width=0.7\linewidth]{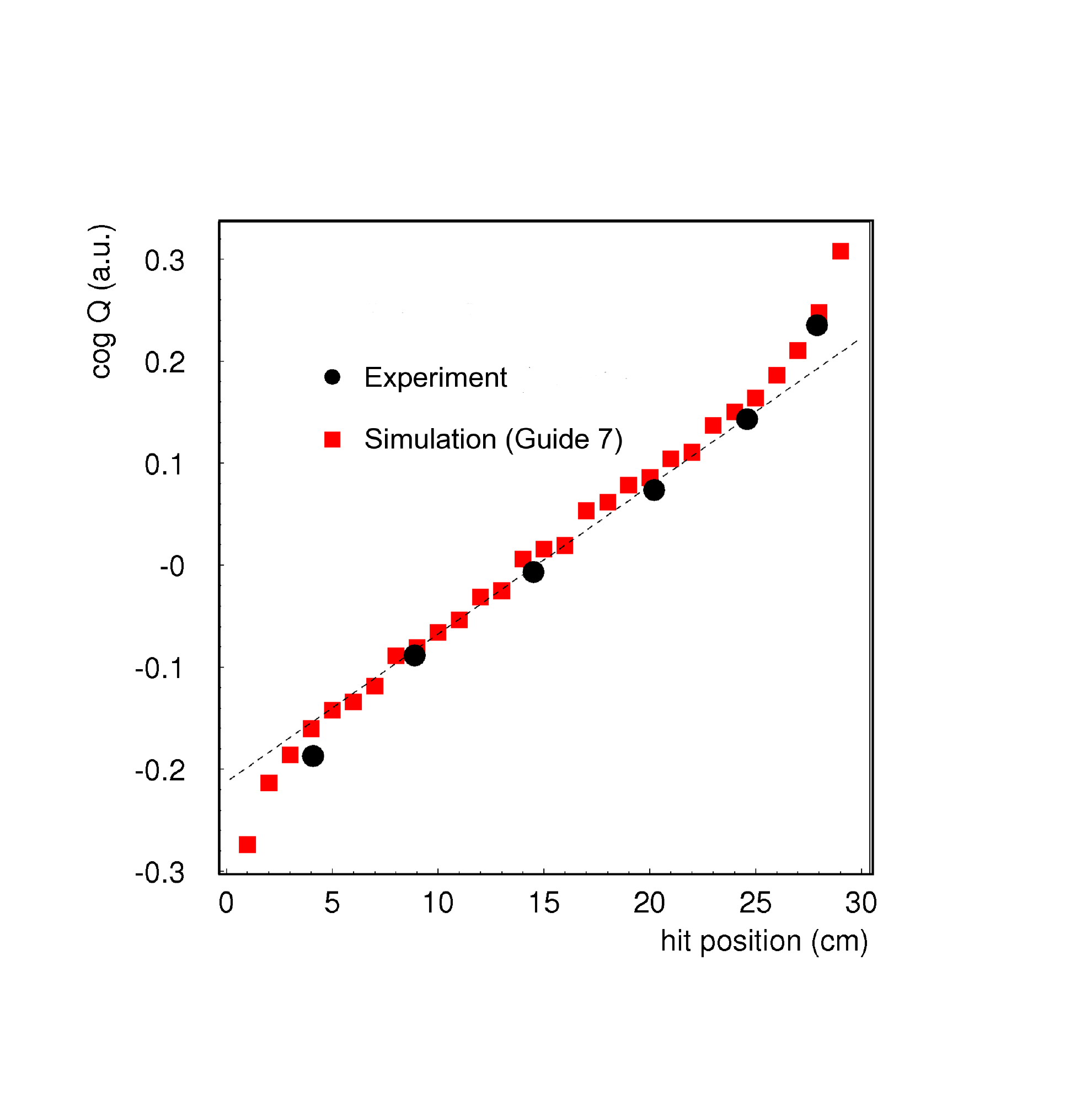}
\caption{Comparison between simulations of the light transport in LENDA with the code GUIDE7 (red squares) and the experimental data of figure \ref{fig:CogvsX} (black dots). The error in the experimental data due to the aperture size is not included explicitly in this graph but the increased size of the experimental points is used to point out the existence of experimental uncertainty. The dashed straight line is a fit excluding the 2 outermost experimental points, used to guide the eye to the deviation from linearity at the edges of the detector.}
\label{fig:guide7_cogq}
\end{figure}

Using the TDC and QDC information from each PMT as described in section \ref{sec:DataReduc}, the position of an event registered by LENDA can be determined. The PMT closer to the event registers the scintillation light sooner and produces a larger signal than the PMT further away. The position resolution therefore depends on the distance traveled by the scintillation light inside the detector as well as the attenuation caused to the scintillation light while traversing the scintillator material and interacting with the reflective coating of the scintillator. The amount of light produced by an event is also a significant factor and the position resolution is expected to be better for higher deposited energies in the scintillator.\\
The position resolution of LENDA bars was determined experimentally using radioactive sources. In earlier investigations using a prototype detector and a $^{252}$Cf source, a clear correlation was observed between the quantities x$_{T}$ and x$_{Q}$ described in equations \ref{eq:x_T} and \ref{eq:x_Q} \cite{Perdikakis09a}. In figure \ref{fig:position_slideshow}, the 2-dimensional spectra of those two parameters are presented for a source placed at different locations along a detector. The clear shift of the locus of events is directly correlated to the change in location of the source.
\\
For the position resolution measurements, a bore hole in a large  lead block was used to collimate $\gamma$ rays from the $^{22}$Na  source. A schematic of the setup is shown in figure \ref{fig:position_setup}. The thickness of the lead ($\approx $10 cm) was enough to block the 511 keV $ \gamma $ rays emitted from $ ^{22} $Na. The diameter of the collimator hole was $1.6$ cm.  The collimator was placed at the minimum distance possible from the detector - 5 cm - to minimize the diameter of the irradiation area. The resulting irradiation spot of 2.4 cm total diameter, was small compared to the position resolution expected by such detectors (several cm FWHM). In this setup, LENDA was moved vertically in front of the collimated source for the measurements. The position of the source center along the bar was read from a scale fixed on the side of the detector.\\
For five different locations of the collimated source, T$_{diff}$ and cogQ (defined above in equations \ref{eq:x_T} and \ref{eq:x_Q} respectively) between the 2 PMTs of each bar were recorded. Typical spectra of these quantities are presented in figure \ref{fig:pos_distributions}. The deviation from a Gaussian shape for the distributions in this figure comes from photons scattering on the support structure. In the analysis, these scattering tails where not considered as valid events and a Gaussian fit was used to determine the centroid values of X$_{Q}$ and X$_{T}$ distributions. 
 The results correlating the position of the source with respect to each one of the two parameters are presented in figures \subref*{fig:TdiffvsX} and \subref*{fig:CogvsX} respectively. No constraint was set on the deposited energy of the events and therefore these results correspond to an average position resolution.  
A linear trend is prominent in both graphs. The small but negligible deviation from linearity observed in the plot of figure \subref*{fig:CogvsX} is  attributed to the presence of edge effects that affect the pulse-height of the signal for events very close to the two extremes of the scintillator. 
To better understand if the transport of light in the scintillator bars can account for this effect, simulations were performed using the code GUIDE7 \cite{Guide7}. The code was modified to correctly reproduce the transport of light in the area between the scintillator material and its wrapping and to include as a parameter the reflectivity of the wrapping material \cite{Cae08a}. 
In figure \ref{fig:guide7_cogq}, the comparison between the simulation results and the experimental data of figure \subref*{fig:CogvsX} are presented. The results of the light transport simulations reproduced fairly well the position dependence observed experimentally.
An average position resolution of 6.3$ \pm $0.2 cm was achieved using the timing information (equation 3) while a slightly worse resolution of 7.2$ \pm $0.3 cm was achieved using the QDC charge signal.
\section{Conclusion}
\label{sec:Conclusion}
LENDA, an array of plastic scintillators capable of detecting low energy neutrons with high efficiency and angular resolution has been built and characterized at NSCL. In this work the results of the characterization of the array using radioactive sources, are presented. The efficiency as well as the timing and position resolutions of LENDA have been determined experimentally. The measurements have been supported by Monte Carlo simulations and general agreement was observed between data and calculations.\\
A low neutron energy threshold of $\sim$150 keV has been demonstrated for LENDA. A timing resolution of $\sim$400 ps and a position resolution of $\sim$6 cm have been obtained. An efficiency greater than 20\% for neutrons below 4 MeV has been measured in the test setup. The results of this work confirm that LENDA is an adequate tool for the study of nuclear reactions in inverse kinematics, especially when low-energy neutron detection is required. Recently, the LENDA array has been used at NSCL to study for the first time the spin-isospin response of the radioactive nucleus $^{56}$Ni \cite{Sas11a}. This was the first experiment with a new measurement technique of (p,n) charge-exchange reactions in inverse kinematics at intermediate energies on unstable isotopes. The results of this investigation proved to be important for improving the description of electron-capture rates on nuclei in the iron region, and for modeling the late evolution of core-collapse and thermonuclear supernovae. 
\section*{Acknowledgments}
This work was supported by the US NSF under Grants PHY-0606007 and PHY-0216783 (JINA).
\section*{References}
\bibliographystyle{model1a-num-names}
\bibliography{LENDA_NIM}
\end{document}